\documentclass[amsfonts,showpacs,tightenlines,aps,12pt,floatfix]{revtex4}
\usepackage{bm}
\usepackage{epsfig}
\usepackage{graphicx}
\begin{document}


\title{A MIXED $\tau$-ELECTROPRODUCTION SUM RULE FOR $\vert V_{us}\vert$}

\author{K. Maltman\footnote{alternate address: CSSM, School of Chemistry
and Physics, University of Adelaide, SA 5005 Australia}}

\affiliation{Department of Mathematics and Statistics, 
York University, 4700 Keele St.,\\
Toronto, Ontario M3J 1P3, Canada}

\date{\today}

\begin{abstract}
The interpretation of results of recent $\tau$ decay determinations of 
$\vert V_{us}\vert$, which yield values $\sim 3\sigma$ low compared to 
3-family unitarity expectations, is complicated by the slow convergence 
of the relevant integrated $D=2$ OPE series. We introduce a class of
new sum rules involving both electroproduction and $\tau$
decay data designed to deal with this problem by strongly suppressing
$D=2$ OPE contributions at the correlator level. 
Experimental complications are briefly discussed and
an example of the improved control over theoretical errors presented.
The uncertainty on the resulting determination, 
$\vert V_{us}\vert =0.2202(39)$, is entirely dominated by experimental
errors, and should be subject to significant near-term improvement.

\end{abstract}


\maketitle

\section{Introduction}	
The CKM matrix element, $\vert V_{us}\vert$, is one of the fundamental
parameters of the Standard Model (SM).
Determinations from multiple sources 
can help to improve the accuracy with which it is known and/or test
for the presence of non-SM contributions in strangeness-changing 
weak processes. Current analyses of $K_{\ell 3}$ and 
$\Gamma [K_{\mu 2}]/\Gamma [\pi_{\mu 2}]$~\cite{marcianovus}, using lattice 
input for $f_+(0)$ and $f_K/f_\pi$, 
respectively~\cite{latticefplus,latticefkfpi}, 
yield values which are in good mutual agreement and
compatible with the expectations of 
3-family unitarity~\cite{hardytowner,juttner07,fnwgk08,kloevus08}.
$\vert V_{us}\vert$ can also be obtained using flavor-breaking (FB) 
hadronic-$\tau$-decay-based sum 
rules~\cite{gamizetal,kmcw,gamizetalnew,mwbrn08}. 
Recent $\tau$ decay analyses~\cite{gamizetalnew,mwbrn08}, employing updated 
information~\cite{babar07a,babar07b,babar08a,babar08b,belle06,belle07a,belle07b} 
on the older strange decay distribution~\cite{alephus99,opalus04},
yield values $\sim 3\sigma$ below 3-family-unitarity expections.

In existing $\tau$ decay determinations, the dominant source of
error on $\vert V_{us}\vert$ is the uncertainties on 
weighted integrals over the inclusive strange decay distribution.
This error will be significantly reduced by ongoing B-factory analyses.
Nominal theoretical errors, estimated with conventional prescriptions,
are small, holding out the prospect of results competitive with those
from $K_{\ell 3}$ and $\Gamma [K_{\mu 2}]/\Gamma [\pi_{\mu 2}]$, 
once the B-factory analyses are complete. A potential complication, however,
arises from the slow convergence of the relevant integrated 
$D=2$ OPE series. Evidence suggests that theoretical errors
may be underestimated (in some cases, significantly) as a 
consequence of this behavior.

In this paper we consider alternate sum rules for $\vert V_{us}\vert$, 
involving both $\tau$ and electroproduction, rather than just $\tau$,
spectral data. The combinations chosen have, by construction, 
{\it already at the correlator level}, a strong suppression of the 
potentially problematic $D=2$ OPE series, and hence also a strongly 
reduced $D=2$ trunctation contribution to the theoretical uncertainty. 
The rest of the paper is organized as follows.
In Section~\ref{sec2} we first briefly outline the purely
$\tau$ decay approach and associated $D=2$ OPE convergence problem. 
Then, in Section~\ref{sec3} we introduce and discuss
the alternate, mixed $\tau$ decay-electroproduction sum rules.
Finally, Section~\ref{sec4} outlines the spectral and OPE input,
discusses some experimental complications, and provides an illustration of
the utility of the mixed sum rule approach.

\section{\label{sec2}The hadronic $\tau$ decay determination of
$\vert V_{us}\vert$}

For any correlator, $\Pi$, without kinematic singularities, and
any analytic weight, $w(s)$, analyticity
implies the finite energy sum rule (FESR) relation,
\begin{equation}
\int_0^{s_0}w(s)\, \rho(s)\, ds\, =\, -{\frac{1}{2\pi i}}\oint_{\vert
s\vert =s_0}w(s)\, \Pi (s)\, ds\ ,
\label{basicfesr}
\end{equation}
where $\rho (s)$ is the spectral function of $\Pi (s)$ and 
the OPE expansion of $\Pi (s)$ can be employed on the RHS
for sufficiently large $s_0$. 
$\vert V_{us}\vert$ is obtained by applying this relation to
the FB correlator difference $\Delta\Pi_\tau (s)\, \equiv\, 
\left[ \Pi_{V+A;ud}^{(0+1)}(s)\, -\, \Pi_{V+A;us}^{(0+1)}(s)\right]$, 
where $\Pi^{(J)}_{V/A;ij}$ are the spin $J=0,1$ components
of the flavor $ij$, vector (V) or axial vector (A) current two-point functions,
and the corresponding spectral functions, $\rho^{(J)}_{V/A;ij}$,
are related to the differential distributions, $dR_{V/A;ij}/ds$, 
of the normalized flavor $ij$ V or A current induced decay widths, 
$R_{V/A;ij}\, \equiv\, \Gamma [\tau^- \rightarrow \nu_\tau
\, {\rm hadrons}_{V/A;ij}\, (\gamma )]/ \Gamma [\tau^- \rightarrow
\nu_\tau e^- {\bar \nu}_e (\gamma)]$, by~\cite{tsai}
\begin{equation}
{\frac{dR_{V/A;ij}}{ds}}\, =\, {\frac{12\pi^2\vert V_{ij}\vert^2 S_{EW}}
{m_\tau^2}}\, \left[ w_T^{(0,0)}(y_\tau )
\rho_{V/A;ij}^{(0+1)}(s) - w_L^{(0,0)}(y_\tau )\rho_{V/A;ij}^{(0)}(s) \right]
\label{basictaudecay}\end{equation}
with $y_\tau =s/m_\tau^2$, $w_T^{(0,0)}(y)=(1-y)^2(1+2y)$,
$w_L^{(0,0)}(y)=2y(1-y)^2$, $V_{ij}$ the flavor $ij$ CKM matrix element, 
$S_{EW}$ a short-distance electroweak correction~\cite{erler02}, and
$(0+1)$ denoting the sum of $J=0$ and $1$ contributions.
The $J=0$ contribution to any $J=0+1/J=0$ decomposition will be referred to
as ``longitudinal'' in what follows.


The use of the $J=0+1$ difference, $\Delta\Pi_\tau$, is the result of
the extremely bad behavior of the integrated longitudinal $D=2$ OPE 
series~\cite{longprob}, which precludes working with FB FESRs based on the 
linear combination of $J=0,0+1$ spectral functions appearing in 
Eq.~(\ref{basictaudecay}). The $\rho^{(0+1)}_{V/A;ij}(s)$, and from these,
$\Delta\rho_\tau$ are obtained by identifying and
subtracting, bin-by-bin, the longitudinal contributions to
$dR_{V/A;ij}/ds$. This can be done with good accuracy because,
apart from the $\pi$ contribution to $\rho^{(0)}_{A;ud}$ and $K$ contribution
to $\rho^{(0)}_{A;us}$ (which are determined by $f_\pi$ and $f_K$, 
respectively, and hence very accurately known) 
all contributions to $\rho^{(0)}_{V/A;ij}$ are doubly chirally 
suppressed, by factors of $O[(m_i\mp m_j)^2]$. The $ij=ud$ longitudinal 
contributions are thus, to high accuracy, saturated by the $\pi$ pole term.
Continuum longitudinal 
$ij=us$ contributions, which are small,
but not entirely negligible, are determined from dispersive~\cite{jop} 
and sum rule~\cite{mksps} analyses of the strange scalar and pseudoscalar 
channels, respectively, analyses which are strongly constrained 
by their implications for $m_s$~\cite{longsubfootnote}.

Given $w(s)$ and $s_0\leq m_\tau^2$, $\vert V_{us}\vert$ is determined by
first constructing, from the longitudinally subtracted
$dR_{V/A;ij}/ds$, the spectral integrals
\begin{equation}
R^w_{V/A;ij}(s_0)\, \equiv\, 12\pi^2 S_{EW} \vert V_{ij}\vert^2\,
\int_0^{s_0}{\frac{ds}{m_\tau^2}}\, w(s)\, \rho^{(0+1)}_{V+A;ij}(s)\ ,
\end{equation}
and, from these, the FB combinations, 
\begin{equation}
\delta R^w_{V+A}(s_0)\, =\, 
\left[ R^w_{V+A;ud}(s_0)/\vert V_{ud}\vert^2\right]
\, -\, \left[ R^w_{V+A;us}(s_0)/\vert V_{us}\vert^2\right]\ .
\label{tauvusbasicidea}\end{equation}
Using the OPE representation of $\delta R_{V+A}^w(s_0)$, and 
inputting $\vert V_{ud}\vert$ and the required OPE parameters
from other sources, one obtains~\cite{gamizetal}, from Eq.~(\ref{basicfesr}),
\begin{equation}
\vert V_{us}\vert \, =\, \sqrt{{\frac{R^w_{V+A;us}(s_0)}
{{\frac{R^w_{V+A;ud}(s_0)}{\vert V_{ud}\vert^2}}
\, -\, \left[\delta R^w_{V+A}(s_0)\right]_{OPE}}}}\ .
\label{tauvussolution}\end{equation}
Since, at scales $s_0\sim 2-3\ {\rm GeV}^2$, 
$\left[\delta R^w_{V/A}(s_0)\right]_{OPE}$ is typically much
smaller than $R^w_{V/A;ud,us}(s_0)$ (usually at the few-to-several-$\%$ 
level), Eq.~(\ref{tauvussolution}) yields a determination of 
$\vert V_{us}\vert$ with a fractional uncertainty {\it much} smaller than that 
on $\left[\delta R^w_{V/A}(s_0)\right]_{OPE}$ itself~\cite{gamizetal}.

A particularly advantageous case, from the point of view of experimental
errors, is that based on $s_0=m_\tau^2$ and the weight 
$w(s)=w_{(00)}(y_\tau )\equiv w_T^{(0,0)}(y_\tau )$
In this case, the $us$ and $ud$ spectral integrals appearing in
Eq.~(\ref{tauvussolution}) are fixed by the total strange and non-strange 
$\tau$ branching fractions, allowing one to take advantage of improvements 
in the errors on a number of the strange branching fractions in advance of
the completion of the remeasurement of the full $us$ spectral distribution.
A disadvantage of this approach is that, working with only a single $s_0$,
one is unable to test the stability of the output $\vert V_{us}\vert$
values with respect to  $s_0$, a crucial step to ensuring that
estimates of the accompanying theoretical uncertainty (which, in some
places in the literature, are quoted to be as low as $0.0005$) are
sufficiently conservative. See below for more on this point.

The OPE representation of $\delta R_{V/A}^w(s_0)$ is, of necessity,
truncated, in both dimension and the perturbative order of the relevant 
Wilson coefficients. Estimating the associated theoretical uncertainty 
is complicated by the less-than-ideal convergence of the $J=0+1$, 
$D=2$ OPE series. Explicitly~\cite{chkw93,bck05}
\begin{equation}
\left[\Delta\Pi_\tau (Q^2)\right]^{OPE}_{D=2}\, =\, {\frac{3}{2\pi^2}}\,
{\frac{m_s(Q^2)}{Q^2}} \left[ 1\, +\, {\frac{7}{3}} \bar{a}\, +\, 
19.93 \bar{a}^2 \, +\, 208.75 \bar{a}^3
\, +\, \cdots \right]\ ,
\label{d2form}\end{equation}
with $\bar{a}=\alpha_s(Q^2)/\pi$, and $\alpha_s(Q^2)$ and $m_s(Q^2)$ 
the running coupling and strange quark mass in the $\overline{MS}$ scheme.
Since independent determinations of 
$\alpha_s$~\cite{bck08,davieretal08,bj08,my08,alphaslatticenew,alphasshapes,alphasothernew,alphasglobalew} 
imply $\bar{a}(m_\tau^2)\simeq 0.10$, the convergence of the $D=2$, 
$J=0+1$ series at the spacelike point on the OPE contour is marginal 
at best. While (at least if one works with the contour improved (CIPT)
prescription~\cite{cipt}, in which the large logs are resummed point-by-point
along the contour) the convergence of the integrated series 
can be improved through appropriate weight choices~\cite{km00}, 
taking into account that $\vert \alpha_s(Q^2)\vert$ decreases as one moves 
away from the spacelike point along the contour, one expects,
in general, rather slow convergence, which makes conventional 
truncation error estimates potentially unreliable. Fortunately, 
the growth of $\alpha_s$ with decreasing $s_0$ means that omitted 
higher-order terms become relatively more important at lower scales, and
hence that any premature truncation of the slowly converging integrated $D=2$, 
$J=0+1$ series will show up as an unphysical $s_0$-dependence in
the extracted values of $\vert V_{us}\vert$. Unphysical $s_0$-dependence
can also be produced by incorrect input for poorly known, or unknown, 
condensates relevant to $D>4$ OPE contributions ($D=6$ and $8$
in the case of $w_{(00)}$).

Such unphysical $s_0$-dependence is, in fact, seen, at a scale significantly 
larger than the estimated $D=2$ truncation error, in recent $\tau$ decay 
analyses~\cite{mwbrn08}. This is illustrated in Figure~\ref{fig1},
which shows results for $w_{(00)}$, and three
additional weights, $w_{10}$, $\hat{w}_{10}$ and $w_{20}$, 
introduced originally to improve the integrated $D=2$, $J=0+1$ 
convergence~\cite{km00}. Of particular note is the situation
for the experimentally favorable $w_{(00)}$ weight case, where the 
instability is much larger than full estimated theoretical error.

It is worth noting that the $D=2$ truncation component of the $0.0005$ 
theoretical error in the $w_{(00)}$ case is obtained by combining an 
uncertainty associated with the residual scale dependence with the shift 
obtained by omitting the last term included in the $D=2$ series, all 
evaluations being performed using the CIPT prescription and the truncated 
$D=2$ Adler function form. Alternate evalutions, using the
truncated correlator (rather than truncated Adler function) form, 
and/or using the fixed order (FOPT) rather than CIPT prescription,
are, however, also possible. At a given, common truncation order,
all such evaluations are equivalent to the CIPT Adler function evaluation, 
differing from it only by corrections of yet-higher order. While the
difference of $\vert V_{us}\vert$ values obtained using the
$O(\bar{a}^3)$ and $O(\bar{a}^4)$ CIPT Adler function evaluations
is, indeed, small ($\delta \vert V_{us}\vert\, =\, -0.0003$),
shifting to alternate $D=2$ evaluation schemes leads to much
larger shifts. For example, shifting from the $O(\bar{a}^3)$ CIPT
Adler function evaluation to the $O(\bar{a}^4)$ CIPT correlator
version yields instead $\delta \vert V_{us}\vert\, =\, -0.0008$,
while shifting from the $O(\bar{a}^4)$ CIPT Adler function (correlator)
versions to the $O(\bar{a}^4)$ FOPT correlator version yields the even larger 
shifts $\delta \vert V_{us}\vert\, =\, 0.0019\, (0.0023)$~\cite{thanksjamin}. 
With plausible arguments in favor of both the CIPT and FOPT prescriptions
in the literature~\cite{bj08,cipt}, such shifts suggest the
conventional $D=2$ truncation error estimate, which leads to
the total estimated theoretical uncertainty, $\delta\vert V_{us}\vert
= 0.0005$, for the $s_0=m_\tau^2$, $w_{(00)}$ determination, 
is far from a conservative one. 

In view of the possibility of much-larger-than-anticipated
$D=2$ truncation uncertainties on the values of $\vert V_{us}\vert$
extracted using the $\Delta\Pi_\tau$ FESRs, we consider, in what
follows, FESRs based on alternate correlator differences designed to have,
already at the correlator level, much reduced $D=2$ contributions.
Such FESRs also allow one to investigate whether the sizeable 
$s_0$-instability observed in the results of the $w_{(00)}$-weighted 
$\Delta\Pi_\tau$ analysis is a consequence of $D=2$ truncation uncertainties, 
or of unexpectedly large $D=6,8$ OPE contributions.

\section{\label{sec3}New mixed $\tau$-electroproduction 
sumrules for $\vert V_{us}\vert$}

Problems associated with the slow convergence 
of the integrated $D=2$, $J=0+1$ OPE series can be reduced by considering
alternate FESRs based on correlator differences, $\Delta\Pi$,
sharing with $\Delta\Pi_\tau$ the vanishing of $D=0$ OPE contributions, 
but having $D=2$ contributions suppressed at the correlator level. 
Since a V/A separation of the flavor $us$ decay distribution
is not presently feasible, 
$\Delta\Pi$ should involve the $us$ V+A combination. The leading order term
in the $D=2$ Wilson coefficient can be removed
by forming the appropriate difference of $\Pi_{V+A;us}^{(0+1)}$
and the electromagnetic (EM) correlator, $\Pi_{EM}$. The following
combinations (having the same $\Pi_{V+A;us}^{(0+1)}$
contribution as $\Delta\Pi_\tau$) have, in addition, vanishing
$D=0$ contributions:
\begin{equation}
\Delta\Pi_\kappa\equiv 9\Pi_{EM}\, -\, \Pi_{V+A;us}^{(0+1)}
\, -\, 2(2+\kappa )\Pi_{V;ud}^{(0+1)}\, 
+\, 2\kappa\Pi_{A;ud}^{(0+1)}\ .
\label{mixedfesrform}\end{equation}
The $\kappa = 1/2$ combination is strictly FB. Bearing in mind
that $\bar{a}(m_\tau^2)\simeq 0.1$, the corresponding $D=2$ OPE contribution,
\begin{equation}
\left[\Delta\Pi_\kappa (Q^2)\right]^{OPE}_{D=2}\, =\, {\frac{3}{2\pi^2}}\,
{\frac{m_s(Q^2)}{Q^2}} \left[ {\frac{1}{3}} \bar{a}\, +\,
4.3839 \bar{a}^2 \, +\, 44.943 \bar{a}^3 \, +\, \cdots \right]
\label{d2altform}\end{equation}
is seen to be strongly suppressed, by more than an order of magnitude,
compared to $\left[\Delta\Pi_\tau\right]_{OPE}^{D=2}$.
A similar suppression turns out to be operative for the $D=4$ contributions.
Explicitly, up to numerically tiny $O(m_s^4)$ corrections, and neglecting,
for simplicity of presentation, $r=(m_d-m_u)/(m_d+m_u)$, one has,
to $O(\bar{a}^2)$~\cite{chkw93,bnp},
\begin{eqnarray}
\left[\Delta\Pi_\tau (Q^2)\right]^{OPE}_{D=4}
&=& {\frac{2}{Q^4}}\left[ \langle m_\ell \bar{\ell}\ell\rangle
\, -\, \langle m_s\bar{s}s\rangle\right]\left( 1-\bar{a}
-{\frac{13}{3}}\bar{a}^2\right)\label{d4form}\\
\left[\Delta\Pi_\kappa (Q^2)\right]^{OPE}_{D=4}
&=&{\frac{2}{Q^4}}\left[ \left(\left({\frac{4-16\kappa}{3}}\right)\bar{a}
+\left({\frac{59-236\kappa}{6}}\right)\bar{a}^2\right) 
\langle m_\ell\bar{\ell}\ell\rangle \right.\nonumber\\
&&\left. \ \ +\left({\frac{4}{3}}\bar{a}
+{\frac{59}{6}}\bar{a}^2\right)\langle m_s\bar{s}s\rangle\right]
\label{d4altform}\end{eqnarray}
where in both cases the strange condensate term is numerically dominant. 

Defining
$R_{EM}^w(s_0)=[12\pi^2 S_{EW}/m_\tau^2]\, \int_0^{s_0}ds\, w(s)\rho_{EM}(s)$ 
and $\left[\delta R^w_\kappa (s_0)\right]^{OPE}
= [12\pi^2 S_{EW}/m_\tau^2]\, \oint_{\vert s\vert = s_0}
ds\, w(s)\left[\Delta\Pi_\kappa (s)\right]^{OPE}$, one then has,
for any analytic $w(s)$ and any $s_0$ large enough the OPE representation 
is reliable, the $\Delta\Pi_\kappa$ analogue of Eq.~(\ref{tauvussolution}),
\begin{equation}
\vert V_{us}\vert \, =\, \sqrt{{\frac{R^w_{V+A;us}(s_0)}
{9R_{EM}^w(s_0)- \left({\frac{2(2+\kappa )
R^w_{V;ud}(s_0)-2\kappa R^w_{A;ud}}{\vert V_{ud}\vert^2}}\right)
\, -\, \left[\Delta R^w_\kappa (s_0)\right]^{OPE}}}}\ .
\label{tauemvussolution}\end{equation}

The suppression of the $D=2$ and $4$ contributions in
$\left[\Delta\Pi_\kappa\right]^{OPE}$ does not persist to higher $D$.
For example, with
$r_c=\langle \bar{s}s\rangle /\langle\bar{\ell}\ell\rangle$,
the $D=6$ contributions, in the vacuum saturation
approximation (VSA), become~\cite{bnp}
\begin{eqnarray}
\left[\Delta\Pi_\tau (Q^2)\right]^{OPE}_{D=6; VSA}
&=& {\frac{\pi\alpha_s \langle \bar{\ell}\ell\rangle^2}{Q^6}}\left[
{\frac{64}{81}}\left( 1-r_c^2\right)\right]\label{d6tauvsa}\\
\left[\Delta\Pi_\kappa (Q^2)\right]^{OPE}_{D=6; VSA}
&=& {\frac{\pi\alpha_s \langle \bar{\ell}\ell\rangle^2}{Q^6}}\left[
\left({\frac{-32+128\kappa}{9}}\right) - {\frac{32r_c^2}{9}}\right]
\label{d6kappavsa}
\end{eqnarray}
typically significantly larger for $\Delta\Pi_\kappa$ than for
$\Delta\Pi_\tau$. 

To deal with such potentially enhanced, but phenomenologically poorly 
determined, $D>4$ contributions, it is useful to 
employ polynomial weights, $w(y)=\sum_{m=0}b_my^m$, with $y=s/s_0$. 
Integrated $D=2k+2$ OPE contributions then scale as $1/s_0^k$. 
The strong suppression of $D=2,4$ contributions, which scale more 
slowly with $s_0$, then means one can, for example, employ the VSA 
estimate for $D=6$, and ignore $D>6$ contributions, but look for
$s_0$ values large enough that $\vert V_{us}\vert$
becomes stable with respect to $s_0$, indicating that $D>4$
contributions and/or deviations from the input assumptions about
these contributions have decreased to a negligible level.

\begin{figure*}
  \begin{minipage}[t]{0.48\linewidth}
\rotatebox{270}{\mbox{
\includegraphics[width=0.9\textwidth]{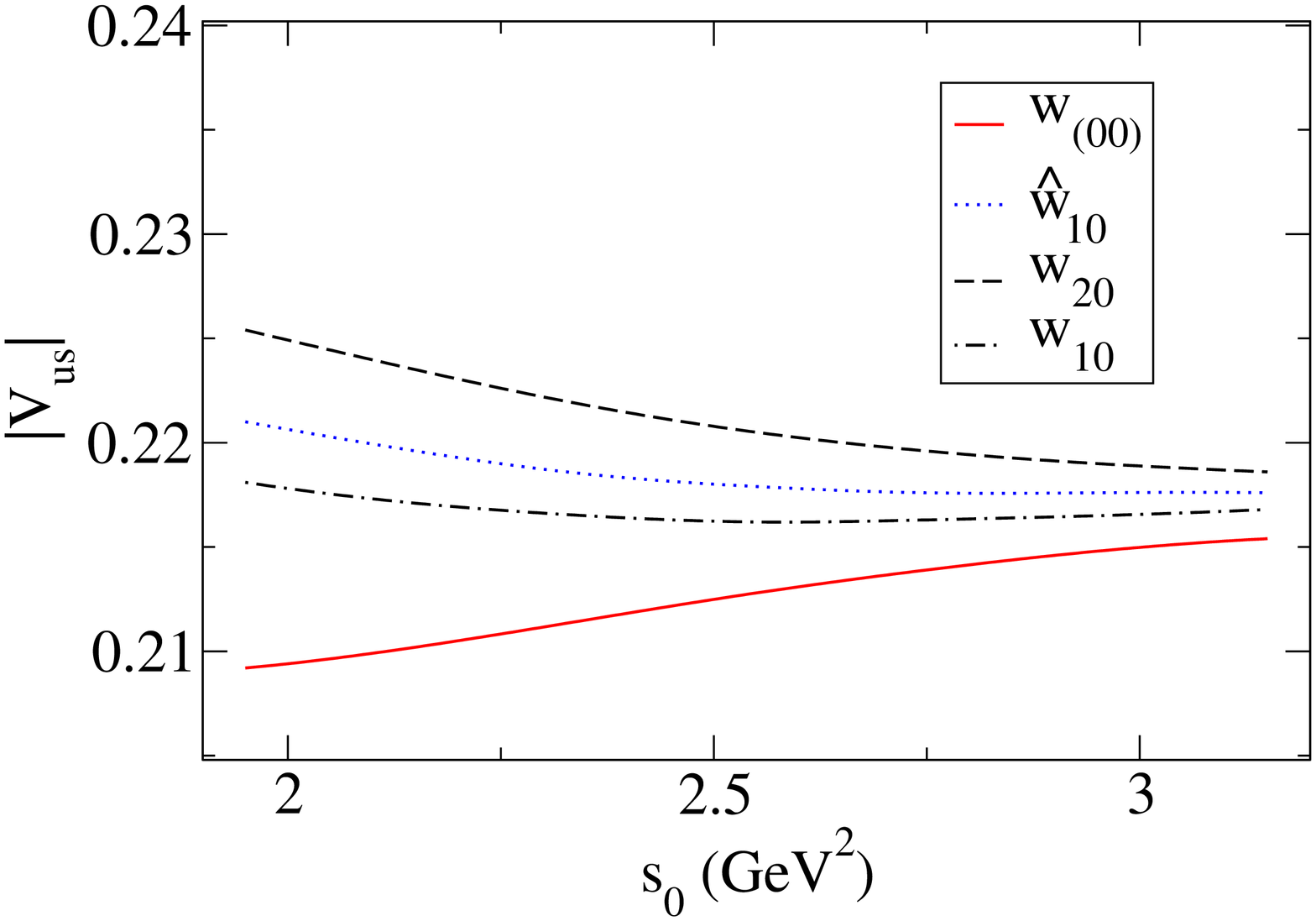}
}}
\caption{$\vert V_{us}\vert$ versus $s_0$ from the $\Delta\Pi_\tau$
FESRs for $w_{20}$, $\hat{w}_{10}$, $w_{10}$ and $w_{(00)}$. \label{fig1}}
  \end{minipage}
\hfill
  \begin{minipage}[t]{0.48\linewidth}
\rotatebox{270}{\mbox{
\includegraphics[width=0.9\textwidth]{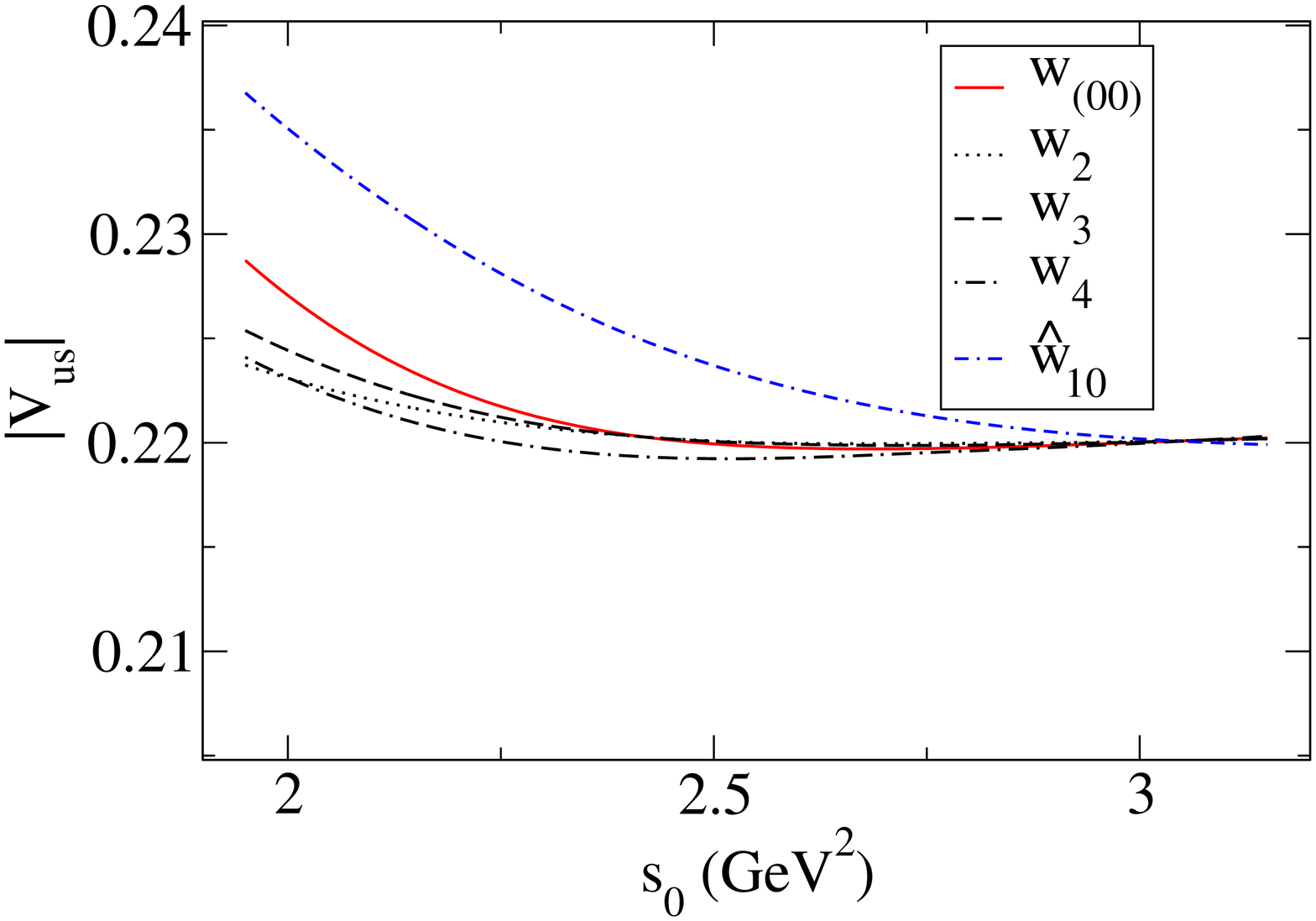}
}}
   \caption{$\vert V_{us}\vert$ versus $s_0$ from the $\Delta\Pi_{\kappa =1/2}$
FESRs for $w_{(00)}$, $\hat{w}_{10}$, $w_2$, $w_3$ and $w_4$. \label{fig2}}
\end{minipage}
\end{figure*}

The expected enhanced role of $D=6$ and higher contributions
in $\Delta\Pi_\kappa$ means that higher degree weights
like $w_{10}$, $w_{20}$ and $\hat{w}_{10}$, introduced to improve
the integrated $D=2$ convergence for the $\Delta\Pi_\tau$ FESRs, are 
likely to represent less useful choices for the $\Delta\Pi_\kappa$ 
analysis. The strong suppression of $D=2$ contributions, 
however, opens up the possibility of using weights which provide less good 
integrated $D=2$ convergence but better control 
over integrated $D>4$ contributions. Thus, e.g., if it is the
slow $D=2$ convergence which is responsible for
the significant $s_0$-instability of the $w_{(00)}$-weighted 
$\Delta\Pi_\tau$ FESR results shown in Figure~\ref{fig1},
the analogous $\Delta\Pi_\kappa$ FESR might be rendered stable 
by the reduced $D=2$ contributions, allowing
improvements in the strange branching fractions
(whose sum provides an improved determination of 
$R^{w_{(00)}}_{V+A;us}(m_\tau^2)$) to be used in reducing the
error on the numerator in Eq.~(\ref{tauemvussolution})
for $w= w_{(00)}$ and $s_0=m_\tau^2$. Similarly, it might become
possible to employ the weights, $w_N(y)=1-{\frac{N}{N-1}}y+{\frac{y^N}{N-1}}$,
which, like $w_{(00)}$, display slow integrated $D=2$
convergence for $\Delta\Pi_\tau$ but are useful for handling $D>4$ 
contributions (written generically as $\sum_{D=6,8,\cdots} C_D/Q^D$) since 
(up to corrections of $O([\alpha_s(m_\tau^2)]^2)$) only a single 
integrated $D>4$ contribution, $(-1)^N C_{2N+2}/[(N-1)s_0^N]$,
survives on the OPE side of the $w_N$ FESR.

\section{\label{sec4}Input, complications, results, and discussion}
In this section we illustrate the utility of the new mixed 
FESRs and point out some experimental complications, focussing on 
the $\Delta\Pi_{\kappa =1/2}$ case, whose $D>4$ contributions vanish
in the $SU(3)_F$ limit, and are thus expected to be optimally suppressed.

\subsection{OPE input}
To suppress OPE-breaking contributions from the region of the contour 
on the RHS of Eq.~(\ref{basicfesr}) near the timelike point
on the contour, we restrict
our attention to $w(s)$ having a zero of order $\geq 2$ at $s=s_0$
and to $s_0>2\ {\rm GeV}^2$~\cite{opebreakdown}.

$D=2$ OPE integrals are evaluated using Eq.~(\ref{d2form}) and the CIPT 
prescription~\cite{cipt}, with $\alpha_s(Q^2)$ and $m_s(Q^2)$ 
the exact solutions associated with the 4-loop-truncated 
$\beta$ and $\gamma$ functions~\cite{4loopbetagamma} and the initial 
conditions, $m_s(m_\tau^2)=100\pm 10$ MeV~\cite{gamizetalnew}, 
$\alpha_s(m_\tau^2)=0.323(17)$. The latter is obtained by running
a very conservative assessment, $0.1190(20)$, of the average of 
several recent independent $\alpha_s(M_Z^2)$
determinations~\cite{bck08,davieretal08,bj08,my08,alphaslatticenew,alphasshapes,alphasothernew,alphasglobalew}
down to the $\tau$ scale using the standard self-consistent combination of 
4-loop running and 3-loop matching at the flavor thresholds~\cite{cks97}.
To be conservative, we assign the sum of absolute values of
the contributions of all computed orders as the truncation component of the
$D=2$ uncertainty (producing a $100\%$ uncertainty if all contributions
have the same sign, larger otherwise). The error on the truncated
$D=2$ sum associated with that on the overall
$\left[ m_s(m_\tau^2)\right]^2$ factor is also evaluated using the conservative
all-absolute-values prescription. The truncation and $m_s^2$-scale
errors are combined in quadrature with the much smaller error induced
by the uncertainty on $\alpha_s(m_\tau^2)$ to obtain the full $D=2$ error.

$D=4$ OPE input and uncertainties are as follows.
$\langle m_\ell\bar{\ell}\ell\rangle$ is fixed using the GMOR relation,
$\langle m_s\bar{s}s\rangle$ using conventional ChPT quark mass 
ratios~\cite{leutwylermq} and the value,
$r_c=\langle \bar{s}s\rangle /\langle \bar{\ell}\ell\rangle =1.2\pm 0.3$ 
obtained by updating the quenched-lattice-data-based determination, 
$r_c=0.8(3)$, of Ref.~\cite{jaminssoverll}
using the average, $f_{B_s}/f_B=1.21(4)$~\cite{latticefbsoverfbave}, of 
recent $n_f=2+1$ lattice determinations~\cite{latticefbsoverfb}.
The strange condensate term dominates both the $D=4$ contribution and its 
error, but produces only a very small impact on $\vert V_{us}\vert$
as a consequnce of the suppression of the coefficient
function seen in Eq.~(\ref{d4altform}).

$D>4$ contributions involve poorly known or phenomenologically undetermined
condensate combinations. We estimate $D=6$ contributions using the VSA and 
ignore $D\geq 8$ contributions. If $D>4$ contributions are small, 
the details of
these assumptions are irrelevant. If not, and the assumptions
are inaccurate, the $1/s_0^2$ ($1/s_0^3,\cdots $) dependence of 
integrated $D=6$ ($8,\cdots$) contributions will lead to
an unphysical $s_0$-dependence of $\vert V_{us}\vert$.
We look for weights which produce a good window of $s_0$-stability
in order to ensure that $D>4$ contributions are either negligible
or estimated with sufficient accuracy.

\subsection{Spectral input}
Results for $R_{V/A;ud}^w(s_0)$ and $R^w_{V+A;us}(s_0)$ are based on 
the ALEPH $us$~\cite{alephus99} and $ud$~\cite{alephud05} spectral data,
for which information on the relevant covariance matrices is publicly 
available. The $ud$ data has been modified to incorporate the recent
improved V/A separation for the $\bar{K}K\pi$ mode~\cite{davieretal08} 
made possible by the BaBar determination of the $I=1$ $K\bar{K}\pi$ 
electroproduction cross-sections~\cite{babarkkbarpi07}. A small rescaling is
applied to the continuum $ud$ V+A distribution to reflect changes in 
$S_{EW}$, $R_{V+A;us}$ and the electron branching fraction, $B_e$. With the 
lepton-universality-constrained result $B_e=0.17818(32)$~\cite{banerjee07} 
and an updated total strange branching fraction $B_{us}=0.02858(71)$, the 
$ud$ normalization is $R_{V+A;ud}=3.478(11)$. For $\vert V_{ud}\vert$,
we use the latest update, $0.97425(23)$, from the super-allowed nuclear 
$0^+\rightarrow 0^+$ $\beta$ decay analysis~\cite{hardytowner}.

Though BaBar and 
Belle have not completed their re-measurements of the inclusive $us$
distribution, $dR_{V+A;us}/ds$, an interim partial update can be obtained 
(following Ref.~\cite{alephrescale}) by rescaling the 1999 ALEPH 
distribution~\cite{alephus99}, mode-by-mode, by the ratio of new to old 
branching fractions. The new branching fraction results are taken from
Refs.~\cite{babar07a,babar07b,babar08a,babar08b,belle06,belle07a,belle07b}.
Unfortunately, this strategy does not allow the corresponding covariance matrix
to be updated. The improved precision on the new strange branching fractions 
can thus be translated into a correspondingly improved $us$ spectral 
integral error only for $w=w_{(00)}$ and $s_0=m_\tau^2$.
Since the recently measured $K$ and $\pi$ branching fractions~\cite{babar08b} 
are compatible with SM expectations at the $\sim 2\sigma$ level, we evaluate 
the $\pi$ and $K$ pole spectral integral contributions using
the more precisely determined $\pi_{\mu 2}$ and $K_{\mu 2}$ input.

$R_{EM}^w(s_0)$ is obtained from the EM spectral function, $\rho_{EM}(s)$,
which is related to the bare $e^+e^-\rightarrow hadrons$ cross-sections, 
$\sigma_{bare}(s)$, by
\begin{equation}
\rho_{EM}(s)\, =\, s\, \sigma_{bare}(s)/16\pi^3\alpha_{EM} (0)^2\ .
\label{rhoemcrosssectionreln}\end{equation}
It is well known that problems exist with the compatibility of the 
measured $\pi\pi$ and $\pi^+\pi^-\pi^0\pi^0$ cross-sections and
those implied by $I=1$ $\tau$ decay
data, even after known isospin-breaking corrections are taken
into account~\cite{dehz}. Preliminary BaBar $\pi^+\pi^-\pi^0\pi^0$ 
cross-section results~\cite{babar4piem} reduce considerably
the latter discrepancy, but have not yet been finalized.
The situation for $\pi\pi$ is somewhat muddier. The recent
KLOE update~\cite{kloepipi08} yields results now in reasonable 
agreement with CMD2 and SND below the $\rho$ peak and with a reduced 
discrepancy above it, while preliminary BaBar
results~\cite{babartau08empipi} are instead in better agreement with $\tau$ 
expectations. In addition, the most recent $\tau$-based analysis~\cite{my08} 
produces an $\alpha_s(M_Z)$ in excellent agreement with 
two recent high-precision lattice determinations~\cite{alphaslatticenew}, 
while electroproduction-based analyses (albeit not updated for new
post-2005 experimental results, and without the careful fitting of 
$D>4$ OPE contributions performed for the $\tau$ case) yield values 
$\sim 2\sigma$ too low~\cite{km05}, again favoring the $\tau$ version
of the $I=1$ spectral distribution. We thus deal with the $I=1$ discrepancies
by replacing both $\pi\pi$ and $4\pi$ EM results with the corresponding
$\tau$ expectations. Since (i) the V/A separation for the $\bar{K}K\pi$
contribution to $\tau$ decay has been performed using CVC and the BaBar 
$I=1$ EM cross-sections and (ii) the $\pi\pi$, $4\pi$ and $\bar{K}K\pi$ 
contributions largely saturate $\rho_{V;ud}^{(0+1)}(s)$ below $s=m_\tau^2$,
this is effectively equivalent to replacing $\Delta\Pi_{1/2}$ with
the alternate combination
\begin{equation}
{\frac{3}{2}}\Pi_{V;I=0} -{\frac{1}{2}}\Pi_{V;ud}^{(0+1)}
+\Pi_{A;ud}^{(0+1)}-\Pi_{V+A;us}^{(0+1)}\ ,
\label{altcorrelator}\end{equation}
where $\Pi_{V;I=0}$ is the $I=0$ octet analogue of $\Pi_{V;ud}^{(0+1)}$.
EM cross-sections are taken from Whalley's 2003 
compilation~\cite{whalley03} and recent updates reported in 
Refs.~\cite{sndnewem,cmd2newem,babarnewem}. Where needed, vacuum
polarization corrections are computed using F. Jegerlehner's
code~\cite{jegerlehnervp}.

\subsection{Results and discussion}

The results for $\vert V_{us}\vert$ as a function of $s_0$
obtained from the $\Delta\Pi_{1/2}$ FESRs for $w_{(00)}$,
$w_2$, $w_3$, $w_4$ and the weight, $\hat{w}_{10}$, producing 
the best $\Delta\Pi_\tau$ $s_0$-stability plateau in Figure~\ref{fig1}
are displayed in Figure~\ref{fig2}. 
In all but the last case a very good $s_0$-stability
plateau is found. In addition, the $\vert V_{us}\vert$
obtained at the highest accessible scale, $s_0=m_\tau^2$ (the right
endpoints of the curves) are all, without exception, in
extremely good agreement. The very good stability plateau for 
$w_{(00)}$ strongly suggests that the instability seen in
the analogous $\Delta\Pi_\tau$ analysis was a result of the slow
$D=2$ convergence. In contrast, the quality of the stability
plateau for $\hat{w}_{10}$ has deteriorated in going from
$\Delta\Pi_\tau$ to $\Delta\Pi_{1/2}$, most likely due to the increased size 
of $D>4$ contributions. Even so, the $\vert V_{us}\vert$ values
for $\hat{w}_{10}$ converge nicely to the stable results from the other
weight cases as $s_0\rightarrow m_\tau^2$.

Because of the very good stability found for $w_{(00)}$,
it is possible to quote a final determination
based on $w=w_{(00)}$ and $s_0=m_\tau^2$, a choice
which allows us to benefit from the improved BaBar and Belle
strange branching fraction determinations. We find
\begin{equation}
\vert V_{us}\vert = 0.2202(27)_{us}(28)_{EM}(2)_{V;ud}(4)_{A;ud}(2)_{OPE}
=0.2202(39)
\label{finalresult}\end{equation}
where the errors are those associated with the inclusive $us$ branching
fraction, the residual $I=0$ EM spectral integral, the residual 
inclusive $ud$ V and $ud$ A branching fractions, and the combined $D=2$ 
and $D=4$ OPE contribution, respectively.

While, within current errors, the result for $\vert V_{us}\vert$ is compatible
with either 3-family-unitarity or the recent $\Delta\Pi_\tau$ 
determinations, and hence does not help in resolving the
$\sim 3\sigma$ discrepancy between the two, prospects exist for siginificantly
reducing the main components of the error. First, 
errors on the weighted $I=0$ EM integrals will be reduced through 
ongoing work on the exclusive EM cross-sections at VEPP2000, BaBar and Belle.
Second, errors on the $us$ spectral integrals will
be significantly reduced by BaBar and Belle analyses
of both the branching fractions of as-yet-unremeasured
strange modes (including the sizable $\bar{K}^0\pi^0\pi^-$ and previously 
estimated, but unmeasured, $\bar{K}3\pi$ and $\bar{K}4\pi$ modes) and
the inclusive $us$ V+A distribution. Obtaining the inclusive $us$ 
distribution, and not just the branching fractions, is crucial to
performing the $s_0$-stability checks, themselves crucial
to demonstrating that $D=2$ convergence and $D>4$ contributions have, indeed,
been brought under good control. To reduce
the $us$-distribution-induced contribution to the error on
$\vert V_{us}\vert$ to, e.g., the $\sim 0.0005$ level requires
$\sim 1.3\times 10^{-4}$ precision on the inclusive $us$
branching fraction, and hence, almost certainly, pursuing
previously undetected higher multiplicity modes 
having branching fractions down to the few $\times 10^{-5}$ level.

We close by stressing the complementarity of the $\Delta\Pi_\tau$ and 
$\Delta\Pi_{1/2}$ analyses. The latter, by construction, has 
significantly reduced OPE-induced uncertainties. The smallness of the 
OPE contributions to the denominator of Eq.~(\ref{tauvussolution}),
however, means that global normalization uncertainties common to the 
$ud$ and $us$ spectral distributions cancel, essentially entirely, in 
the $\Delta\Pi_\tau$ determination. This is not the case for the 
$\Delta\Pi_{1/2}$ analysis, where EM and $\tau$ normalization uncertainties
are independent, leading to an increased experimental error on 
$\vert V_{us}\vert$. As we have seen already in the $w_{(00)}$
case, employing the same weight in both FESRs and comparing the 
$s_0$-dependences of the resulting $\vert V_{us}\vert$ determinations
can also help in shedding light on the source of any $s_0$-instabilities found
in the $\Delta\Pi_\tau$ analysis, where OPE-induced errors are
more difficult to reliably quantify.

\section{Acknowledgments}

The ongoing support of the Natural Sciences and
Engineering Research Council of Canada, as well as the
hospitality of the Theory Group at IHEP, Beijing and the CSSM
at the University of Adelaide are gratefully acknowledged.
Thanks also to Fred Jegerlehner for providing
his code for computing vacuum polarization corrections.


\begin{thebibliography}{0}
\bibitem{longsubfootnote}As a measure of the accuracy with which 
$\rho_{V/A;ij}^{(0+1)}$ can be determined, note that $\sim 20\%$ shifts 
in the $us$ longitudinal continuum contributions induce shifts of 
$\sim 0.0002$ in $\vert V_{us}\vert$.
\end{thebibliography}
\end{document}